\documentstyle[epsfig,12pt]{article}
\renewcommand{\topfraction}{0.99}
\renewcommand{\bottomfraction}{0.99}
\renewcommand{\textfraction}{0.0}
\renewcommand{\baselinestretch}{1.5}
\renewcommand{\thefootnote}{\fnsymbol{footnote}}

\begin{document}

\newcommand{\3}{\ss}

\makebox[14cm][r]{PITHA 94/36}\par
\makebox[14cm][r]{TTP 94--18} \par
\makebox[14cm][r]{September 1994}\par
\vspace{.7cm}
\centerline{\Large Some Properties of a Transient New Coherent
Condition of Matter}
\par
\centerline{\Large Formed in High--Energy Hadronic Collisions}
\vspace{1.cm}
\centerline{Saul Barshay$^{a)} $ and Patrick Heiliger$^{b)} $ }
\par
\centerline{$^{a)} $III. Physikalisches Institut (A), RWTH Aachen}
\par
\centerline{D--52056 Aachen, Germany}
\centerline{$^{b)} $Institut f\"ur Theoretische Teilchenphysik} \par
\centerline{Universit\"at Karlsruhe}\par
\centerline{D--76128 Karlsruhe, Germany}\par
\normalsize
\vspace{1.cm}

\begin{abstract}
We investigate the dynamical possibility for the formation of a
transient new coherent condition of matter in high--energy hadronic
collisions. The coherent bosonic amplitude is characterized by a
non--zero momentum and is sustained by $P $--wave interactions of
quasi--pions in a dense fermionic medium. We make quantitative
estimates of several essential properties: the condensate momentum
and the fermionic density, the size of the coherent amplitude
and the negative energy density contributed by the condensate, a
characteristic proper time for the system to exist prior to
breakdown into a few pions, and a characteristic extension of the
system over the plane perpendicular to the collision axis. These
quantities then allow us to make definite estimates of new signals:
a few pions with anomalously small transverse momenta $ \leq 50 $
MeV/c; and a possible anomalous bremsstrahlung of very soft photons
with characteristic transverse momenta as low as about 4 MeV/c.
\end{abstract}

\newpage
There exist certain definite anomalous effects in current
high--energy experiments involving collisions of hadrons
\cite{cohmat, chlia, bot, soph1, soph2}. These unusual effects can
be connected through the transient existence of a new coherent
condition of matter \cite{cohmat}. Such a hypothetical, relatively
long--lived system, becoming spatially extended over the
(impact--parameter) plane perpendicular to the collision axis can
be formed in some of the real, intermediate inelastic states which
contribute to the (imaginary) amplitude for the diffractive elastic
scattering of hadrons. The coherent system can support an oscillatory
behavior of matter with a fairly definite non--zero momentum
\cite{arn} in the transverse plane. The relevant Fourier--Bessel
transform of this spatial oscillatory behavior into the space of
transverse momentum--transfer $\sqrt{-t} \simeq {\sqrt{s} \over 2}
\theta $ (for small c.m. scattering angle $\theta $), results in a
localized structure in the differential cross section $\displaystyle
{d\sigma_{el}\over dt} $ around a fairly definite value of $\sqrt{-t
} $. Such a localized structure exists \cite{cohmat} in the data
from collider and fixed--target experiments up to the present
highest center--of--mass energy $\sqrt{s} = 1800 $ GeV. There must
also be direct signals from the transient coherent system in
inelastic processes. One signal that is a general consequence of the
extension of the system to several fermis over the transverse plane
is the breakdown into a few pions with anomalously small transverse
momenta $\leq 50 $ MeV/c. Observation of this unusual effect against
the background of many pions produced with the usual average
transverse momentum of $(250-350) $ MeV/c requires measurements on
individual events, in particular events with low multiplicity. In
fact, this unusual phenomenon of small--$p_T $, few--particle
emission is the main new empirical feature in a recent, detailed
study of several hundred cosmic--ray showers ``in the forwardmost
small--angular region'' \cite{pam}. If charged currents exist within
the coherent system during its evolution, then it has been shown
that soft photons will be emitted \cite{saul}. The occurrence of an
anomalously high emission of soft photons has been and is today, an
enigmatic feature of several different experiments at CERN
\cite{chlia, bot,soph1, soph2, emc}. \par
\bigskip
In this paper we attempt to calculate approximately some properties
of a transient new coherent condition of matter formed in
high--energy hadronic collisions. We utilize general properties of a
physical model in order to obtain numerical estimates for some
definite properties of the hypothetical, new coherent condition.
The model is patterned after the phenomenon of pion condensation
in a sufficiently dense medium of nucleons \cite{migd, brown}. The
pions are replaced by quasi--pions which are correlated
quark--antiquark pairs with small $(\mbox{mass})^2 $. The dense
(``cold'') medium of nucleons is replaced by a dense medium of
``dressing'' quarks and antiquarks formed in the collision. In this
medium the quasi--pions propagate and interact with the fermions via
strong, low--energy $P $--wave interactions, just as pions do in a
medium of nucleons. In this model we examine the possibility for the
occurrence of a coherent condition which contributes a negative
energy density in the extending medium. The essential
\underline{dynamical} element which makes this possible is the
existence of a pseudoscalar (quasi--) boson with quite small
$(\mbox{mass})^2 $ and strong, low--energy $P $--wave interaction
with fermions. This allows for the possibility of cancellation of the
positive, kinetic and $(\mbox{mass})^2 $ terms by the attractive
momentum--dependent interaction \cite{migd, brown}. The hypothetical
coherent condition which is dynamically instigated following the
collision, is evolving toward some (idealized) equilibrium situation,
but it is limited by the intrinsic, though extended proper time for
the system to break down into a few pions. Therefore one property
that we seek a reason for and estimate of is the extended proper time
for the system to exist prior to breaking down. In general, the
center of mass of the developing coherent system can move rapidly
along the collision axis in the collision center of mass (and more
rapidly along the beam direction in the laboratory system). This
longitudinal dimension is Lorentz--contracted; the intrinsic time
interval is dilated.
A second property that we estimate is the extension of the system
over the transverse plane during the intrinsic time interval prior
to its breakdown into a few pions. We estimate the magnitude of the
momentum which characterizes the coherent oscillatory behavior of
matter over the transverse plane, and we estimate the fermionic
density which sustains this behavior. We calculate the negative
energy density contributed by the coherent structure, which depends
upon its squared amplitude. The latter quantity together with the
square of the estimated transverse extension of the system makes
possible an estimate of the amplitude for anomalous soft--photon
emission. In the system, soft photons are emitted with a
bremsstrahlung--like energy spectrum \cite{saul} ( $\propto
1/p_{\gamma } $), at least down to $p_{\gamma } $ of the order of
$1/T $, where $T $ is the intrinsic time interval before breakdown.
In the laboratory the anomalous photons appear in a small angular
interval (very small $p_T $) around the collision axis because of the
assumed rapid motion of the system. The few pions formed when the
system breaks down have anomalously small $p_T $, that is transverse
momenta of the order of $1/R_b $ or less, where $R_b $ is the
transverse extension at $T $. In some events, these pions (in
particular, $\pi^+ \pi^- $ pairs) can be anomalously ``bunched'' in
rapidity. In concluding we discuss possible energy dependence once
$\sqrt{s} $ is such that sufficient fermionic densities are
attainable following the collision; in particular the likelihood that
the coherent system has some probability to break down into a large
number of pions only when created in collisions with $\sqrt{s} $
well up in the TeV range. We also remark the possibility that a meson
beam can be more effective at lower energies than a baryon beam in
instigating the condensate. \par
\bigskip
A high density of bare quarks and antiquarks is likely to occur just
after the high--energy collision of the extended structured hadrons,
with an initial extension of the order of $1/m_{\pi} = 1.4 $ fm over
the impact--parameter plane. These entities start to dress themselves
via QCD self--interactions, and to become correlated toward forming
pions as constituent quarks and antiquarks. In the latter process
it is usually tacitly assumed that interaction signals occur with
nearly the velocity of light over distances of the order of
$1/m_{\pi} $. Therefore the sequential emission of 4 to 8 pions
might involve a proper time of about $4/m_{\pi} - 8/m_{\pi} =
(2-4) \cdot 10^{-23} \; sec $. However as the dynamical degrees of
freedom switch from those of QCD to constituent quarks and to
quasi--pions -- correlated quark--antiquark pairs with small
$(\mbox{mass})^2 $ -- a coherent condition can begin to evolve over
the impact--parameter plane instigated by the strong, low--momentum
$P $--wave interaction of quasi--pions with ``dressing'' quarks and
antiquarks. As we calculate below, the effective signal velocity
in the extending coherent medium is not close to that of light. The
proper time for the breakdown of the system into some physical pions
can be significantly extended. This has direct physical
consequences. Since the center of mass of the coherent system
generally can move with a high speed along the collision axis in the
collision c.m., it will arrive at a longitudinal point rather
distant from the collision point before breakdown, during the
Lorentz--dilated proper time. The entire spatial domain involves
evolution of the coherent system. The longitudinal extent of the
effective interaction region is much greater than is usual in
$p(\overline{p})-p $ and $\pi-p $ collisions, in those inelastic
events in which the system occurs. If charged currents
within the system are emitting bremsstrahlung \cite{saul}, these
photons can have very low energies, a few MeV or down to of order of
the inverse of the above extended proper time, before they cut off
\cite{low}. \par
\bigskip
We examine the possibility of a zero in the following dispersion
relation, some time after the collision when the dynamics is assumed
to have largely become that relevant to quasi--pions with effective
mass $\mu $, interacting coherently with a dense medium of
constituent quarks and antiquarks with effective mass $m $. The
fermionic medium is represented by a Fermi sea with momenta from 0
to $p_F $. If physical conditions for a zero in the dispersion
relation are attained then the amplitude for a coherent boson field
in the medium can emerge. (When appearing as numbers, all energies
and momenta are henceforth to be understood as in units of $\mu $.)
\begin{equation}
\omega^2 = 1 + k^2 - k^2 F^2(k) \Pi (k, p_F) \Longrightarrow 0
\end{equation}
The $(1 + k^2) $ terms are from the mass and kinetic energy; the
negative third term results from the successive attractive $P $--wave
interactions of nearly zero--energy ($\omega \to 0 $) quasi--pions
with the fermionic medium. The essential dynamical element is
contained in the fact that the strong $P $--wave scattering
interaction is attractive and starts out as $k^2 $; we have included
a phenomenological $[ $F 1$] $ form factor $F(k) $ to damp the
interaction strength for $k \geq 3 $. The ``self--energy'' function
$\Pi $, for the pion in the medium is given explicitly by the
following generalization of the Lindhard function \cite{lind}
utilized in solids \cite{schrieff} for the interaction of phonons
with the electron Fermi sea
\begin{eqnarray}
\Pi (k, p_F) = {12\over \pi^2} {f^2\over 4 \pi} & & \!\!\!\!\!\!\!\!
\int_{L_x}^{p_F} 2 \, dx \int_{L_y}^{p_F} y \, dy \int_0^{\overline
\varphi} 2 \; d \varphi  \nonumber \\
& & \!\!\! \cdot {1\over (x^2 + y^2 + k^2 + 2 y k \cos \varphi +
m^2)^{1/2} - (x^2 + y^2 + m^2)^{1/2} } \\
& & \!\!\!\!\!\!\!\!\!\!\!\!\!\!\!
 \mbox{with} \;\;\; {\overline{\varphi}} =
\cos^{-1} \left ( {-k^2 + 2 p_F^2 - y^2 -x^2 \over 2 y k } \right )
\nonumber \\
& & L_y = \sqrt{2 p_F^2 - x^2} -k \nonumber \\
& & L_x = {\cases {  0 \;\;\;\;\;\;\;\;\;\;\;\;\;\;\;\;\;\;\;\;\;\;
                   \;\;\;\;\;\;\;\;\;
                   \mbox{for} \;\; k  > (\sqrt{2}-1) p_F
                   & \cr
                   (2p_F^2 - (p_F+k)^2)^{1/2}  \;\; \mbox{for} \;\;
          k < (\sqrt{2}-1) p_F > 0  } } \nonumber
\end{eqnarray}
We are evaluating the interaction function in cylindrical
coordinates. The coherent interactions take place in the plane
transverse to the collision axis. A quasi--pion with a small
momentum  $\vec{k} $ directed in this plane interacts with a fermion
with transverse--momentum component $\vec{y} $ for $|\vec{y}| < p_F
$, resulting in the intermediate--state fermion with a transverse
momentum $({\vec y} + {\vec k}) $ and a total (squared) momentum
$({\vec x}^2 + |{\vec y} + {\vec k}|^2) > 2 p_F^2 $ ( a ``hole'' in
the Fermi sea of momentum states). Coherence probably cannot develop
with momentum ${\vec k} $ in the longitudinal direction because the
system generally moves with considerable velocity along the
collision axis in the collision c.m. These coherent interactions
tend to be impeded by (relatively small) differences in this
longitudinal motion in the parts of the system. The ``strength'' of
$\Pi $ is essentially governed by the coupling parameter $[ $F 2$] $
$(f^2/4 \pi) \sim 0.08 $, and by $p_F^2 $ which is related to the
fermionic density $[ $F 3$] $ by $\displaystyle \rho = {3 p_F^3\over
\pi^2} \left [0.36 \left ({\mu \over m_{\pi}} \right )^3 fm^{-3}
\right ] $. Thus this density is of the order of $1 \; fm^{-3} $ for
$p_F $ of the order of 2 with $\mu = m_{\pi} $. Alternatively,
$p_F \leq 3.5 $ corresponds to approximately one dressing quark (or
one antiquark) of a given color in a volume of about 1 $(\mbox{fm}
)^3 $. After expanding the denominator in eq. (2) in the
$k $--dependent terms, we have evaluated $\Pi $ with numerical
integration and have examined the zero--condition in eq. (1). With
$[ $F 1$] $ $F(k) = e^{-(k/6.2)} $ there are in fact zeros in the
sensible domains $k < k_0 < p_F $ and $p_F < 3.5 $. In Table 1 we
list several paired values of $k_0, p_F $. Note that $k_0 $
increases as $p_F $ decreases, as is intuitively expected. This is
because in the interaction term in eq. (1) we are in a region of
$k $ where the $k^2 $ factor still controls the strength, rather
than $F^2(k) $. For each fixed $p_F $, we have gradually increased
$k $ until the minimum in the dispersion relation is reached. The
most negative value of $\omega^2 $ is $(\omega^2)_{min} = -
|\omega^2_{min}| $, occurring at $k=k_c $. These quantitites are
also given for the examples in Table 1. \par
\bigskip
The quantity $\omega_{min}^2 $ determines the
negative energy density contributed by the coherent quasi--pion
condition in the fermionic medium according to an estimate carried
out in the following manner. We minimize the condensate energy with
respect to the amplitude $A $ of a trial (static $[ $F4$] $) wave
function of the (running--wave $[ $F5$] $) form $\displaystyle \phi
({\vec r}) = A e^{ik_c x} $ (for $\pi^- $ with $\displaystyle A
e^{-ik_c x} $ for $\pi^+ $; x lies in the transverse plane). For the
expectation value of the Hamiltonian in the condensate we use
\begin{eqnarray}
<H> & = & \left < {1\over 2} \int d^3{\vec r} \phi_i^{\dagger}
(\mu^2 - {\vec \nabla}^2 ) \phi_i \right. \nonumber \\
& & + {1\over 2} \int dk_x dk_y dk_z \{k_x^2 F(k_x^2) \Pi (k_x, p_F)
\} \phi_i^{\dagger} ({\vec k}) \phi_i({\vec k}) \nonumber \\
& & \left. - {1\over 16 F^2} \int d^3 {\vec r} \left \{ 4
(\partial_{\mu} \phi_i^{\dagger} \phi_i ) (\phi_j^{\dagger}
\partial^{\mu} \phi_j) + 2 \mu^2 (\phi_i^{\dagger} \phi_i) (\phi_j
\phi_j^{\dagger}) \right \} \right > \nonumber \\
\mbox{where} & & \phi({\vec k}) = A \delta (k_x-k_c) \delta(k_y)
\delta(k_z)
\end{eqnarray}
The indices $i, j $ are summed over $\pi^+ $ and $\pi^- $ and we have
modelled a (momentum--dependent) quartic interaction between
quasi--pions by the form suggested by chiral dynamics \cite{wein},
with $F \sim (\mu /m_{\pi})\, F_{\pi}$ ($F_{\pi} = 95 $ MeV). This
term gives an effective interaction which is repulsive, and
stabilizes the condensate \cite{migd, brown}. The minimization
condition then reads as follows
\begin{equation}
0 = ( - |(\omega^2)_{min}| + 8 |A|^2 \lambda ) V
\end {equation}
with $\lambda = (4 k_c^2 - 2 \mu^2)/ 16 F^2 $ and $V $ the volume of
the system. The result for $|A|^2 $ is thus
\begin{equation}
|A|^2 = |(\omega^2)_{min}|/ 8 \lambda \sim 0.07 \mu^2
\end{equation}
Substitution of (5) into eq. (3) makes possible an estimate of the
condensate energy density $[ $F 5$] $
\begin{equation}
E_c/V = - |(\omega_{min}^2)|^2/16 \lambda \sim -1.5 {\mbox{MeV} \over
\mbox{fm}^3 }
\end{equation}
In this estimate  we have used $\omega_{min}^2 \simeq - 0.9 $ and
$\lambda \sim 1.7 $ for $k_c \sim 1.9 $, with $\mu \sim m_{\pi} $.
\par
\bigskip
As a check on the approximate stability of the results in Table 1, we
have examined the domains of $k_0, p_F $ for zeros in the dispersion
relation in eq. (1), using the standard analytical form for the
Lindhard function as evaluated in a spherical system
\cite{schrieff}. This amounts to replacing eq. (2) by
\begin{eqnarray}
\Pi_L (k, p_F) & = & {12 \over \pi^2} \left ({f^2 \over 4 \pi}
\right ) (\pi m p_F') \nonumber \\
& & \left [ 1 + {p_F' \over k} \left \{1 - \left ({k \over 2 p_F'}
\right )^2 \right \} ln \left |{1 + \left ({k \over 2 p_F'} \right )
\over 1 - \left ({k \over 2 p_F'} \right ) } \right | \right ]
\end{eqnarray}
with $p_F' $ related to the $p_F $ in the cylindrical generalization
in eq. (2) by $p_F' = \sqrt{2} p_F $. We use $F(k) = e^{-(k/6.2)} $.
In Table 2 we list several paired values of $k_0, p_F' $. Again for
each fixed $p_F' $ we gradually increase $k $ until the minimum in
the dispersion relation is reached. These negative values of
$\omega_{min}^2 $ are given in Table 2 together with the values of
$k = k_c $ at which they occur. Somewhat smaller values of $k_0, p_F'
$ occur than in the cylindrical generalization, but the minimum is
less pronounced. In the cylindrical evaluation we have observed that
the purely phenomenological function $F(k) $ which damps the
interaction at high $k $ cannot be made too weak i.e. like $e^{-(k/B)
} $ with $B > 6.2 $. This is because then a minimum value
$\omega_{min}^2 $ is not readily obtained for fixed $p_F $ with
moderate increase of $k \, (\leq p_F) $, within the present
approximations for $\Pi $ in the dispersion relation eq. (1). Of
course, even more pronounced values of $\omega_{min}^2 $ than those
discussed here imply increased metastability of the hypothetical
coherent system. For the present development we feel that the
essential element is in showing the occurrence of the zeros in the
dispersion relation for sensible domains of $k_0, p_F $; and this
with minimal dependence upon the arbitrary form factor $F(k) $. \par
\bigskip
Consider a situation in the collision when, as an ``initial''
condition, a degree of coherence has been set up in the transverse
plane; assume that the dynamics is near the condition
$\omega_{min}^2 (k_c) $. How does the coherence extend in the plane
during the intrinsic time available before the system breaks down ?
We estimate this behavior by using the following approximate
expression for $\delta \omega (\delta k_x) $ near to $|
\omega_{min}^2 |^{1/2} $ for a change $\delta k_x $ near to $k_x
= k_c $.
\begin{eqnarray}
\omega^2 & \simeq & - |\omega_{min}^2| + {d^2 \omega^2 \over d^2k}
|_{k_c} \; {(\delta k_x)^2 \over 2} \nonumber \\
\Rightarrow \omega (\delta k_x) & = & i \{ |\omega_{min}^2|^{1/2}
- D (\delta k_x)^2 \} = i |\omega_{min}^2|^{1/2} + \delta \omega
(\delta k_x) \nonumber \\
\mbox{with} & & D = {d^2 \omega^2 \over d^2 k}|_{k_c} {1\over 4
|\omega_{min}^2|^{1/2} }
\end{eqnarray}
Formally doing a momentum--space Fourier transform over $u = \delta
k_x $ for large times, using the steepest--descent approximation,
gives
\begin{eqnarray}
& & \int_{- \infty}^{+ \infty} du \; e^{iux} e^{-i \delta \omega (u)
t}\; \Pi \simeq \int_{- \infty}^{+ \infty} du \; e^{iux - D u^2 t} \;
\Pi \nonumber \\
& \simeq & e^{-x^2/4 D t} \cdot \sqrt{{\pi \over D t}} \cdot \Pi
(k_c(x), p_F(x)) \propto  e^{- x^2/4 R_x^2 (t)}
\end{eqnarray}
We have used a local density approximation $(p_f(x)) $ in $\Pi $ and
have taken $u_{sad} = ix/2 D t $. Eq. (9) indicates that the
$x $--extension of the interaction is in part controlled by a
function of time $R_x = \sqrt{Dt} $: The extension has the
possibility to become large only as $\sqrt{t} $ where large $t $ is
taken from the ``initial'' collision condition, and $x $ is added to
the initial transverse extension in the collision, $\sim 1/m_{\pi} $.
An effective velocity for the increase of $R_x $ may be defined from
\begin{eqnarray}
R_x & = & v(t) t \nonumber \\
\Rightarrow v(t) & = & \sqrt{D/t}
\end{eqnarray}
An average velocity during an ``apriori'' proper time $T $ for the
system is thus
\begin{equation}
{\overline v} (T) = 1/T \int_0^T dt \; v(t) = 2 \sqrt{{D\over T}}
\end{equation}
This velocity generally is significantly less than unity (as
estimated below). If viewed as an upper limit for signal propagation
as the dynamical degrees of freedom switch from those of QCD to
those of the coherent interactions, it implies an extension of the
proper time interval prior to breakdown, to of the order of $T_p
\sim T/{\overline v} $. In this extended intrinsic time interval,
a characteristic transverse extension is then
\begin{equation}
{\overline R}_x (T_p) = 1/T_p \int_0^{T_p} dt \; R_x(t) = {2\over 3}
\sqrt{D T_p} \simeq {1\over 3} (4D)^{1/4} (T)^{3/4}
\end{equation}
The above argument is heuristic. The physical points are probably
relevant for such an evolving dynamical system. Since $4 D = {d^2
\omega^2 \over d^2 k}|_{k_c} / |(\omega_{min}^2)|^{1/2} $ is of order
unity $[ $F 6$] $ (in unit $1/ \mu $) one can readily get a feeling
for the numbers: for $T \simeq 8/m_{\pi} = 4 \cdot 10^{-23} \; sec $,
${\overline v}(T) \sim 0.35, \; T_p \sim 22.5 /m_{\pi}, \; {\overline
R}_x(T_p) \sim 1.6 /m_{\pi} $ (for $\mu \sim m_{\pi} $). Adding the
initial $\sim 1/m_{\pi} $ to ${\overline R}_x $ gives $ 2.6/m_{\pi}
$. This number and that for $T_p $ have immediate physical
consequences \cite{cohmat}, $[ $F 7$] $. (1) The breakdown of the
coherent system results in a few pions with anomalously small
transverse momenta $\sim 1/ {\overline R}_x \simeq 50 $ MeV/c. (2)
Very soft photons can be emitted in the system, down to energies of
the order of $1/T_p \simeq 6 $ MeV, appearing with laboratory
transverse momenta $\leq 4 $ MeV/c. \par
\bigskip
We turn now to a more detailed discussion \cite{saul} of the possible
emission of soft photons by a hypothetical, new coherent condition
of matter formed in high--energy collisions \cite{cohmat}. We assume
that charged currents exist during the evolution of the system prior
to its breakdown into some pions. The relevant approximate
amplitudes are given by $A e^{ik_0 x} $ for $\pi^- $ and $A
e^{-ik_0 x} $ for $\pi^+ $, with values for $A $ and $k_0 $ that we
have estimated. Using also our estimate for $T_p $ and ${\overline
R}_x(T_p) $, we calculate the soft--photon energy spectrum and
estimate the emission strength.
\par \bigskip
In the coherent system, the matrix element \cite{saul} for a charged
quasi--pion to propagate to a point in the medium, to emit there a
very soft photon with momentum ${\vec p}_{\gamma} $ and polarization
$\hat{\epsilon}_{\gamma} $, and to propagate further in the medium,
is
\begin{eqnarray}
|{\cal M}_{\gamma}| & = &
\Bigg | \left (2 e \left [(2 {\overline R}_x)A \right ]^2 \right )
\; (2 {\vec k}_0 \cdot {\hat \epsilon}_{\gamma} ) \Big / \nonumber \\
& & \left \{ \left ( \omega - \left [\mu^2 + \left ( {\vec k}_0 +
{{\vec p}_{\gamma} \over 2} \right )^2 + {\overline \Pi} \left (
\omega, {\vec k}_0 + {{\vec p}_{\gamma} \over 2}, p_F \right )
\right ]^{1/2} \right ) \right. \nonumber \\
& & \left. \cdot  \left ( \omega - \left [\mu^2 + \left ( {\vec k}_0
 - {{\vec p}_{\gamma} \over 2} \right )^2 + {\overline \Pi} \left (
\omega, {\vec k}_0 - {{\vec p}_{\gamma} \over 2}, p_F \right )
\right ]^{1/2} \right )
\right \} \Bigg | \nonumber \\
& \stackrel{\simeq}{\omega, p_{\gamma} \Rightarrow 0} &
\;\; \Bigg | { \left (2e \left [(2 {\overline R}_x) A \right ]^2
\right ) (1/c) \left (2 {\vec k}_0 \cdot {\vec \epsilon}_{\gamma}
\right ) \over \left ( \left ( {\vec p}_{\gamma} \cdot {\vec k}_0
\right )^2 + \left ({Im \; {\overline \Pi} \over c }\right )^2
\right )^{1/2} } \Bigg |
\end{eqnarray}
where in terms of the $\Pi $ defined by eqs. (1, 2), we have here
${\overline \Pi} (k, p_F) = k^2 F^2(k) \newline \cdot \Pi (k, p_F)
$, and $|c| = |1 + { {\overline \Pi} \over k_0^2} | \sim {\mu^2
\over k_0^2} $. We have evaluated $|{\cal M}_{\gamma} |
$ at $ \omega^2 (k_0) \Rightarrow 0 $, the limit of a usual
perturbative treatment with oscillatory behavior of amplitudes in
time. Continuation to $\omega^2 < 0 $ involves that in $|A|^2
\propto |\omega^2| \neq 0 $. In eq. (13) we have retained the
leading terms as $p_{\gamma} $ formally goes to zero. The
approximate strength (in units of charge $e $) is evident in the
overall factor $[(2 {\overline R}_x) A]^2 $ which arises at the
vertex for bremsstrahlung from the charged--pion condensate
$[ $F 8$] $; the additional factor of 2 occurs from adding
coherently the contribution of oppositely--directed $\pi^- $
and $\pi^+ $ current filaments $[ $F 9$] $. Our previous estimates
$A \sim 0.26 \mu $ and $2 {\overline R}_x \sim (5/ \mu) $, allow an
immediate feeling for the significant probability factor $E $
\begin{equation}
E \sim \left ( 2 \left [ (2 {\overline R}_x) A \right ]^2 \right )^2
\simeq 12
\end{equation}
In eq. (13), we have utilized a necessary, phenomenological
imaginary--part addition to ${\overline \Pi} $. We write
\begin{eqnarray}
Im \; {\overline \Pi} & = & \left \{ \left ( Im \; {\overline \Pi}_s
\right )^2 + \left ( Im \; {\overline \Pi}_t \right )^2 \right
\}^{1/2} \nonumber \\
Im \; {\overline \Pi_t} & = & (\mu/T_p) \sim m_{\pi}^2/22.5
\nonumber \\
Im \; {\overline \Pi_s} & \simeq & \left ( 1/\pi R_x^2 (t) \right )
\simeq 4u/ \pi t \sim 4 m_{\pi} p_{\gamma} / \pi
\end{eqnarray}
The essential lower limit \cite{low} on the anomalous
$p_{\gamma} $--spectrum is incorporated through the characteristic
intrinsic (long) time $T_p $ for the ultimate breakdown of the
coherent system into some pions; this is in $Im \; {\overline \Pi}_t
$, and to be definite we have used our numerical estimate for $T_p $
and put $\mu = m_{\pi} $. There is another physical effect which
inhibits development of the condensate over a time interval $t $
following the instigation of a coherent condition: this is
scattering interactions, with small momentum transfer, in the
fermionic density over the transverse plane. This effect is
incorporated through $Im \; {\overline \Pi_s} $, where we have
assumed the probability for such scatterings to be approximately
proportional to the inverse of the effective transverse area at
$t $. We have used eq. (10) giving $R_x^2 (t) \propto t $, and have
set $1/t \sim p_{\gamma} $ as a measure of the progressive smallness
of the relevant momenta--transfer as time increases. From eqs.
(13--15) we calculate the differential probability for anomalous
soft--photon bremsstrahlung in the phase--space element
$d\rho_{\gamma} $ where, for simplicity, we have used a polar angle
$\theta $ of ${\vec p}_{\gamma} $ with respect to ${\vec k}_c $
(a direction in the transverse plane in each event).
\begin{eqnarray}
|{\cal M}_{\gamma}|^2 \, d\rho_{\gamma} & = & (E/c^2) (e^2/2 \pi^2)
(dp_{\gamma} /p_{\gamma} ) (d\phi/2 \pi) d (\cos \theta) \nonumber \\
& & \cdot {k_0^2 \sin^2 \theta \over \left [ k_0^2 \cos^2 \theta +
{ \left \{ (Im \; {\overline \Pi}_s)^2 + (Im \; {\overline \Pi}_t)^2
\right \} \over c^2 p_{\gamma}^2 } \right ] }
\end{eqnarray}
The $p_{\gamma} $ spectrum is bremsstrahlung--like \cite{saul}. It is
nearly proportional to $ (1/p_{\gamma}) $ until it is cut--off by
the part of the denominator involving $Im \; {\overline \Pi}_t
\propto 1/T_p $. For smaller $p_{\gamma} $ the probability approaches
zero as $p_{\gamma} $ in accord with a basic theorem \cite{low}. The
cut--off clearly occurs for $p_{\gamma} $ so low that the following
condition holds,
\begin{eqnarray}
& & Im \; {\overline \Pi}_s/p_{\gamma} < Im \; {\overline \Pi}_t/
p_{\gamma} \nonumber \\
& & \Rightarrow p_{\gamma} < (\pi/4) \; (m_{\pi}/22.5) \sim 5 \;
{\mbox MeV}
\end{eqnarray}
This value for $p_{\gamma} $ based upon our definite numerical
estimate in eq. (15) is low on the usual hadronic scale of the order
of $m_{\pi}/2 = 70 $ MeV. The anomalous photons are quite soft.
Integrating over the angles in eq. (16) gives the differential
$p_{\gamma} $ probability, with the overall weight--factor $({E K
\over c^2}) \geq 1 $.
\begin{eqnarray}
{d P(\gamma) \over dp_{\gamma} } & = & \left ( {E K \over c^2}
\right ) \left ({e^2 \over \pi^2} \right ) \left ( {1\over p_{\gamma}
}\right ) \nonumber \\
\mbox{where} \; \; K & = & \left \{ {(1 + B) \over B^{1/2} }
\right \} \tan^{-1} \left ({1\over B^{1/2}} \right ) - 1 \nonumber \\
\mbox{with} \;\; B & = & {1\over c^2} {1\over k_0^2 p_{\gamma}^2}
\left \{ (Im \; {\overline \Pi}_s)^2 + (Im \; {\overline \Pi}_t)^2
\right \}
\end{eqnarray}
Where could these soft photons appear in the laboratory system, say
in an experiment with 280 GeV $\pi^- $ on a proton target
\cite{soph1, soph2} ? Since the $\phi $ distribution in eq. (16) is
uniform (this azimuthal angle is in a plane perpendicular to the
transverse plane), in the coherent system the photons have comparable
components of momentum along the collision axis and perpendicular to
it. An immediate consequence is the possibility of observing
anomalous photons with exceedingly low transverse momenta in the
laboratory; a characteristic value being of the order of $(5/\sqrt{2}
) $ MeV/c. We assume \cite{cohmat} that the center of mass of the
coherent system tends to move rapidly along the collision axis in the
collision c.m., in particular along the $\pi^- $--beam direction.
(Note our reason for the $\pi^- $ direction stated in the
next--to--the--last paragraph of the paper.) In addition there is
the motion of the collision c.m. with respect to the laboratory
system. With an overall $\gamma(v) = (1 - v^2)^{-1/2} \sim 280 $,
the low--$p_T $ photons appear inside a cone of about 10
milliradians around the beam direction, with laboratory energies of
a few hundred MeV. Their center--of--mass--rapidities are between
about +2 and +4. \par
\bigskip
It is clear that apart from a possible, conceptually important
soft--photon signal, a signal from events with a few pions that have
anomalously small transverse momenta ($\leq 50 $ MeV/c) should occur.
Anomalous soft--photon events would be correlated with these; it is
possible that photon emission instigates the breakdown. The estimate
that the coherent system typically breaks down into few pions is
simply based upon the estimate of its transverse extent, $(2
{\overline R}_x) \sim 5/m_{\pi} $. This dimension accomodates 2 or
4 pions (not overlapping), with $2<r_{\pi}^2>^{1/2} \sim 1/m_{\pi} $
$[ $F 10$] $. The observational problem is that there may often be
some pions with the usual characteristic transverse momentum of
about 300 MeV/c also present in the events; therefore the average
will be only somewhat lower. One must look at the \underline
{distribution} of pion transverse momenta in individual events;
preferably those with low multiplicity (compared to $<n(s)> $). It
is noteworthy that the recent, comprehensive study of several
hundred individual cosmic--ray events \cite{pam} has concluded that
there exists a class of showers that are highly collimated along the
projectile direction and that carry a large energy fraction with a
penetrating, hadronic character. The authors emphasize \cite{pam}
the unusual feature that these showers seem to often involve only
few particles (hadrons and photons) at production, but with
anomalously small transverse momenta, of the order of 10 MeV/c. \par
\bigskip
It seems likely that once $\sqrt{s} $ is high enough to produce high
densities (in a particular collision $A + B $) that allow instigation
of the coherent effect and which, although falling, can sustain it
for a time, the growth with energy is moderate. For example in $p(
{\overline p})-p $ collisions, it may grow as the central opacity
which is a measure of an initial, interacting--matter density
induced by the collision of fully--overlapping hadrons. According to
a good parameterization \cite{we} the opacity in head--on $p(
{\overline p})-p $ collisions grows by about $40 \% $ between
$\sqrt{s} =53 $ and 546 GeV, and by another $20 \% $ between 546
and 1800 GeV. The complete inelastic cross section grows \cite{we}
in about the same way percentage--wise; thus formation of the
coherent system occurs as approximately a fixed fraction of
$\sigma_{inel} $ (but a somewhat decreasing fraction of
$\sigma_{total} $). It is possible that an energetic boson
projectile ($\pi, K $) is more effective at lower $\sqrt{s} $ in
instigating the quasi-pion condensate. Such a boson projectile
clearly carries much energy into a potential collision involving
(quasi-)bosons $[ $F 11$] $; this is generally not the case for
$p({\overline p})-p $ collisions, or $\sqrt{s} $ must be higher.
The dynamical circumstance which might become more probable at
$\sqrt{s} $ in the multi--TeV range is the breakdown into a large
number of pions ($\geq <n(s)> $). This is possible if $T_p $ and
$\overline{R}_x (T_p) $ become significantly larger \cite{cohmat}.
If this were to happen, then events could occur with the proverbial
\cite{pam} large number of $\pi^- \simeq \pi^+ $ and few or no
$\pi^0 $, or vice versa, since the coherent system is either
principally \cite{world} a charged condensate $(\pi^-, \pi^+) $, or
principally a $\pi^0 $ condensate. In general, the relative
probability for such an ``unmistakable burst'' is low, and thus this
feature of the system is difficult to use for establishing its
presence \cite{ua5, jac,bj} $[ $F 12$] $. \par
\bigskip
This paper has given dynamical reasons for the possibility of the
transient formation of a new coherent condition of matter in
high--energy hadronic collisions. We have quantitatively estimated a
number of properties of the system and have enumerated their related
observational consequences for current experiments. It is possible
that there is already evidence for the formation of such a new
condition \cite{cohmat, chlia, bot, soph1, soph2, pam, emc}. This is
a physical idea which must be guided by experiments.

\newpage
{\large \bf Appendix} \par

\bigskip
We give in this appendix a summary of the origin of the
phenomenological formula which we have recently used to represent
the anomalous behavior in the diffractive, elastic scattering of
hadrons \cite{cohmat}. The derivation is carried out within the
general geometric framework for high--energy scattering \cite{glaub}.
The specific dynamical parameters utilized \cite{cohmat} are related
to those estimated in the present paper to be characteristic of the
new coherent condition. We consider the occurrence of this system
in some of the real, inelastic intermediate states which contribute
to the amplitude for diffractive scattering, as the dynamical origin
of anomalous structure in the amplitude. \par
\bigskip
The structure of Glauber's formula \cite{glaub} for the imaginary,
high--energy amplitude ${\overline F}(\vec{k}', \vec{k}) $, for
scattering  from initial momentum $\vec{k} $ to final $\vec{k}' $, is
given in terms of the eikonal $\Omega $, by
\begin{eqnarray}
{\overline F}(\vec{k}', \vec{k}) & = & {F(\vec{k}', \vec{k}) \over
k} \nonumber \\
& = & i \int {d^2 \vec{b} \over 2 \pi} \int_{-\infty}^{+\infty} dz
e^{-i \vec{k}' \cdot (\vec{b} + \vec{z})} \Omega (\vec{b} + \vec{z})
\left \{ e^{i \vec{k} \cdot (\vec{b} + \vec{z}) - \int_{-\infty}^z
dz' \Omega (\vec{b} + \vec{z}') } \right \} \;\;\;\;\;\;\;\;\; (A 1)
\nonumber
\end{eqnarray}
The derivation \cite{glaub} of the form of the wave function
embodied by the factor in brackets $\{.\,.\,.\} $, involves the
restriction to sufficiently small (c.m.) scattering angles $\theta $
such that $\{{k \theta^2 \over 2}\} z_m < 1 $, where the bracketed
quantity is the approximated longitudinal momentum transfer ($ \simeq
(-t)/\sqrt{s} $) and $z_m $ is a characteristic, maximal longitudinal
dimension within which the primary collision dynamics occurs (as
parameterized by $\Omega $). Without this constraint, additional
phase factors occur in the exponentiated quantity. Consistent with
the same constraint, one usually replaces the factor $e^{i (\vec{k}
- \vec{k}') \cdot \vec{z} } $ by unity; this brings the $z $
integration to that of an exact differential, with the textbook
result \cite{glaub},
\newpage
\begin{eqnarray}
\;\;\;\;\;\;\;\;
{\overline F}(\vec{k}', \vec{k}) & = & i \int_0^{\infty} db \;b \,
J_0 (\sqrt{-t} \, b)\,(1 - e^{-\Omega(b,s)}) \nonumber \\
\;\;\;\;\;\;\;\;
\mbox{with} \; \; \Omega(b,s) & = & \int_{-\infty}^{+\infty} dz' \;
\Omega(\vec{b} + \vec{z}\,'), \;\; \mbox{and} \;\; -t = (\vec{k}' -
\vec{k})^2 \simeq \left ({\sqrt{s}\over 2} \, \theta \right )^2
\nonumber \\
& & \;\;\;\;\;\;\;\;\;\;\;\;\;\;\;\;\;\;\;\;\;\;\;\;\;\;\;\;\;\;\;\;
\;\;\;\;\;\;\;\;\; k' = k = {\sqrt{s} \over 2} \;\;\;\;\;\;\;\;\;\;\;
\;\;\;\;\;\;\;\;\;\;\;\; (A 2)
\nonumber
\end{eqnarray}
We treat, perturbatively, formation of the condensate as an
occasional, dynamical consequence of the initial stage of the
collision, and add phenomenologically $\{\Delta \Omega \} $ to the
eikonal. We neglect this addition in the bracketed quantity in
eq. $(A1) $, and extend the upper limit of the $z' $ integration to
$z' = \infty $. The principal contribution in this integration domain
occurs around $z' = 0 $ since the primary, overlapping configuration
of interacting partons involved in the collision is strongly Lorentz
contracted along the collision axis at high energies. The addition
to the diffractive amplitude is then
$$ \;\;\;\;\;\;\;
\Delta {\overline F} (\vec{k}', \vec{k}, \vec{k}_c) \simeq i
\int {d^2 \vec{b} \over 2 \pi} \int_{-\infty}^{\infty} dz e^{-i
{\vec k}' \cdot (\vec{b} + \vec{z}) } \{ \Delta \Omega (\vec{b} +
\vec{z}) \} e^{i (\vec{k} - \vec{k}_c) \cdot (\vec{b} + \vec{z})}
e^{-\Omega (b)} \;\;\;\;\;\;\;\;\; (A 3)
$$
In eq. $(A3) $, the momentum $(\vec{k} - \vec{k_c}) $ which
characterizes the ``incoming wave--function'' is the initial
(collision--c.m.) momentum reduced by the momentum $\vec{k}_c $
associated with formation of the condensate where $\vec{k}_c = \{
\sqrt{-t_c}\, \hat{b} + ({-t_c \over \sqrt{s}})\, \hat{z} \} $ and
is thus largely transverse. The magnitude $|\vec{k} - \vec{k}_c |$
is effectively the momentum of the coherent system, including
initial projectile. An oscillatory behavior in the transverse plane
with a small wave number $k_c $ appears through the factor $e^{-i
\vec{k}_c \cdot \vec{b}} $. We alternatively use $ \{\Delta \Omega
(\vec{b} + \vec{z}) \}\, e^{-i \vec{k}_c \cdot \vec{b}} \Rightarrow
\{f(z) f(b) \} J_0(k_c b) $ in eq. $(A3) $, leading to
\begin{eqnarray}
\;\;\;\;\;\;\;\;
\Delta {\overline F} & \simeq & i \int_{-\infty}^{\infty} dz \,
e^{i \{(\vec{k}' - \vec{k}) - \vec{k}_c \} \cdot \vec{z}} \, f(z) \,
\int_0^{\infty} db \; b\, J_0(y b)\, J_0(y_c b)\, f(b) \, e^{-
\Omega (b)} \nonumber \\
\;\;\;\;\;\;\;\;\;
\mbox{where} & & \;\; y = \sqrt{-t}, \;\; y_c = \sqrt{-t_c}
 \;\;\;\;\;\;\;\;\;\;\;\;\;\;\;\;\;\;\;\;\;\;\;\;\;\;\;\;\;\;\;
\;\;\;\;\;\;\;\;\;\;\;\;\;\;\;\;\;\; (A4)
\nonumber
\end{eqnarray}
Here we do not replace the exponential in $z $ by unity because the
oscillations in this factor can be relevant over the extended
interaction--domain of $z $ in which the condensate formation takes
place. This occurs out to some distant $z_h $ where the coherent
system breaks down into a few pions after a relatively long proper
time (which is Lorentz dilated). \par
\bigskip
Using as a first approximation (condensate formation is favored in
central collisions where $\Omega (b \sim 0) $ is large),
$$
\{f(z) f(b)\} \simeq [\delta (z - z_h) + \delta(z + z_h)] (N \,
e^{\Omega (b)}) $$
in eq. $(A4) $, gives
\begin{eqnarray}
\;\;\;\;\;
\Delta {\overline F}(t,s) & = & i \left [2 \cos \left \{(y^2 -
y_c^2) \left ({T_h \over 2 m(s)} \right ) \right \} \right ]
\left (N {\delta (y - y_c) \over y_c} \right ) \nonumber \\
\;\;\;\;\;
& \sim & i N  \left [ \cos \left \{(y^2 - y_c^2) \left ({T_h \over
2 m(s)} \right ) \right \} \right ] \left ({R_x \over \sqrt{\pi}
y_c} \, e^{- (y - y_c)^2 R_x^2/4} \right ) \;\;\;\;\;\;\;\;\;\; (A 5)
\nonumber
\end{eqnarray}
Generally a sum of weighted $\delta $--functions could represent
different points $z_h' < z_h $, with the largest weight at $z_h $
where the condensate is most fully developed just before breakdown;
$z_h $ relates to the most rapid oscillatory behavior in $\Delta
{\overline F} $. For a more realistic phenomenological form
\cite{cohmat} of the local structure near $y = y_c $, we have
smeared the $\delta $--function maintaining the overall
normalization (in general $y_c $ fluctuates on an event basis). In
the argument of the oscillatory function the longitudinal
momentum--transfer is $\simeq y^2/\sqrt{s}  \simeq \sqrt{s}
\theta^2/4 $ and the Lorentz--dilated time interval (leading to $z_h
$) is $(\sqrt{s} T_h/ 2m(s)) $, where $m(s) $ is an effective
mass--like parameter which includes $1/\lambda(s) $ where $\lambda(s)
$ is the fraction of $\sqrt{s}/2 $ involved in the (assumed) rapid
motion of the center of mass of the coherent system in the collision
c.m. The parameters $y_c, T_h, R_x $ have been estimated in this
paper $[ $F 13$] $. The size of the anomaly is controlled by the
dimensionless $N \simeq 2 |A|^2 {\overline \sigma} $, where $|A|^2 $
has been estimated in eq. (5) of this paper and ${\overline{\sigma}
} $ is a characteristic cross section associated with the initial
stage of condensate formation in a particular collision system $A +
B $; empirically \cite{cohmat} $\overline{\sigma} \sim 1 $ mb.

\newpage
\large
\noindent
{\bf Footnotes}
\normalsize
\begin{itemize}
\item[[F 1]] This purely phenomenological form factor includes, in
             addition to the cut--off of the interaction of
             quasi--pions with dressing quarks at high $k $, the
             diminishing amplitude for a quark and antiquark to be
             correlated as a quasi--pion with high overall momentum
             $k $.
\item[[F 2]] In our numerical work we use an estimate for this
             coupling parameter given by the number relevant for the
             low--energy, $P $--wave pion--nucleon interaction. There
             $f_{\pi N}^2/{4 \pi} = (m_{\pi}/{2 m_N})^2
             (g_{\pi N}^2/{4 \pi}) $, where $ g_{\pi N}^2/ 4 \pi
             \sim 14.5 $ is the usual pseudoscalar coupling. In
             the present case $f^2/4 \pi = (\mu/2m)^2 (g^2/4 \pi) $;
             we use $(\mu/m)\, g \sim (m_{\pi}/m_N)\, g_{\pi N} $,
             and $m \sim 2.5 \mu $ in eq. (2).
\item[[F 3]] We use the weight of two spin orientations and
             three colors for quarks and antiquarks.
\item[[F 4]] As we have stated, the system is evolving in time up to
             the time of breakdown. This treatment is an
             approximation in a time interval following the onset of
             condensate structure as the breakdown time is
             approached.
\item[[F 5]] The standing--wave form $A \sin k_x x $ which could
             give a more negative condensate energy density may
             evolve later, in particular for a ``$\pi^0 $''
             condensate.
\item[[F 6]] For the examples listed in Table 1, the approximate
             value is $4D \sim 2 $.
\item[[F 7]] In \cite{cohmat} we have utilized these numerical
             estimates in a phenomenological representation of the
             localized structure in the differential cross section
             for diffractive, elastic scattering.
\item[[F 8]] Formally in eq. (3), $\delta (k_x - k_c) \Rightarrow
             (2 {\overline R}_x) $ for a system whose transverse
             extension is not arbitrarily large.
\item[[F 9]] In this paper we do not address the attractive,
             magnetic interaction--energy in such configurations.
             This energy appears to favor the $\pi^-- \pi^+ $
             condensate over the $\pi^0 $ condensate.
\item[[F 10]] The variable effective mass of the system includes the
             negative condensate energy; thus it seems likely to be
             spread from near $2 m_{\pi} $ to below $\sim 4 m_{\pi}
             $. In a sense, the system appears like a variable--mass
             coherent excitation of a ``dressing'' constituent quark.
\item[[F 11]] If such a bosonic collision is involved at the first
             stage, the transverse deflection of the incident pion
             may sometimes determine the direction of the condensate
             momentum in individual events.
\item[[F 12]] Of course observation of one such event in
             highly--controlled circumstances \cite{bj} may be
             enough. Reference \cite{bj} seeks a signal from a
             hypothetical, transient change in the ``vacuum''. A
             dynamical reason for instigation of this change is
             absent. In the present paper we have discussed a
             coherent dynamical structure of \underline{matter} which
             gives rise to a non--zero expectation value of a
             pion--like field.
\item[[F 13]] The parameter $y_c $ is approximately twice the
             condensate momentum calculated in the text (i.e. $2k_0
             < y_c < 2 k_c $), because the oscillatory behavior of
             matter in the transverse plane involves the square of
             the condensate amplitude: $\sim |A \, e^{ik_c x} + A
             e^{-ik_c x}|^2 = 2 |A|^2 \{1 + \cos{2k_c x} \} $. The
             first term is incorporated in the mean distribution and
             the second term represents the matter oscillations
             associated with the condensate.
\end{itemize}


\newpage
\begin{thebibliography}{99}
%
\bibitem{cohmat} S. Barshay and P. Heiliger, ``A Transient New
                Coherent Condition of Matter: The Signal for New
                Physics in Hadronic Diffractive Scattering'',
                Aachen, Karls\-ruhe preprint PITHA 94/15, TTP 94--03,
                March 1994, appears in Z. Phys. C 64 (1994).
\bibitem{chlia} P. V. Chliapnikov et al., Phys. Lett. B 141 (1984)
                276.
\bibitem{bot}   NA 22 Collab., F. Botterweck et al., Z. Phys. C 51
                (1991) 54.
\bibitem{soph1} SOPHIE/WA 83 Collab., S. Banerjee et al., Phys.
                Lett. B 305 (1993) 182.
\bibitem{soph2} SOPHIE/WA 83 Collab., ``Some Aspects of Anomalous
                Soft Photon Production in $\pi-p $ Interactions at
                280 GeV'', Report by T. J. Brodbeck (1993).
\bibitem{arn}   S. Barshay and R. Arnold, Nuovo Cim. 35 A (1976)
                457.
\bibitem{pam}   Chacaltaya and Pamir Collab., L. T. Baradzei et al.,
                Nucl. Phys. B 370 (1992) 365.
\bibitem{saul}  S. Barshay, Phys. Lett. B 227 (1989) 274.
\bibitem{emc}   EMC Collab., J. J. Aubert et al., Phys. Lett. B 218
                (1989) 258.
\bibitem{migd}  A. B. Migdal, Sov. Phys. JETP 34 (1972) 1184; 36
                (1973) 1052.
\bibitem{brown} S. Barshay, G. E. Brown and G. Vagradov, Phys. Lett.
                B 43 (1973) 359.
\bibitem{low}   F. E. Low, Phys. Rev. 110 (1958) 974.
\bibitem{lind}  J. Lindhard, Kgl. Danske Videnskab, Selskab, Mat.
                Fys. Medd. 28 (1954) 8.
\bibitem{schrieff} J. R. Schrieffer, ``Theory of Superconductivity'',
                W. A. Benjamin, Inc., Publishers, New York, Amsterdam
                1964, p. 139.
\bibitem{wein}  S. Weinberg, Comm. in Nuclear and Particle Physics
                2 (1968) 28.
\bibitem{we}   S. Barshay, P. Heiliger and D. Rein, Z. Phys. C 56
                (1992) 77.
\bibitem{world} S. Barshay, Particle World 3 (1994) 180.
\bibitem{ua5}   UA1 Collab., G. Arnison et al., Phys. Lett. B 122
                (1983) 189; \newline
                UA5 Collab., K. Alpgard et al., Phys. Lett. B 115
                (1982) 71; \newline
                UA5 Collab., G. H. Alner et al., Phys. Lett. B 180
                (1986) 415.
\bibitem{jac}   JACEE Collab., J. Lord and J. Awai, paper 515, Int.
                Conf. on HEP, Dallas, Texas, Aug. 1992; J. Awai,
                Univ. of Washington preprint UWSEA 92--06.
\bibitem{bj}    Minimax proposal T 864, J. Bjorken and C. Taylor,
                Fermilab, April 1993.
\bibitem{glaub} R. J. Glauber, Lectures on theoretical physics,
                Vol. 1, p. 315. W. E. Brittin et al. (eds.), New
                York, Interscience 1959.
\end{thebibliography}
\end{document}